\documentclass[conference]{IEEEtran}
\IEEEoverridecommandlockouts
\usepackage{cite}
\usepackage{amsmath,amssymb,amsfonts}
\usepackage{algorithmic}
\usepackage{graphicx}
\usepackage{textcomp}
\usepackage{wasysym}      
\usepackage{fontawesome5} 
\usepackage{booktabs}
\usepackage{multirow}
\usepackage{makecell}   
\usepackage{graphicx}   
\usepackage{xcolor}
\usepackage{hyperref}
\def\BibTeX{{\rm B\kern-.05em{\sc i\kern-.025em b}\kern-.08em
    T\kern-.1667em\lower.7ex\hbox{E}\kern-.125emX}}

\newcommand{\BR}{BookReconciler \faBook\ \faGem}

\begin{document}

\title{BookReconciler \faBook\ \faGem:\\
An Open-Source Tool for Metadata Enrichment\\ and Work-Level Clustering}


\author{\IEEEauthorblockN{Matthew Miller}
\IEEEauthorblockA{
\textit{Library of Congress}\\
Washington, DC, USA\\
thisismattmiller@gmail.com}
\and
\IEEEauthorblockN{Dan Sinykin}
\IEEEauthorblockA{
\textit{Emory University}\\
Atlanta, Georgia, USA \\
daniel.sinykin@emory.edu} 
\and
\IEEEauthorblockN{Melanie Walsh}
\IEEEauthorblockA{\textit{ University of Washington Information School} \\
Seattle, Washington, USA \\
\href{https://orcid.org/0000-0003-4558-3310}{0000-0003-4558-3310}}  
}

\maketitle

\begin{abstract}
We present \BR, an open-source tool for enhancing and clustering book data. \BR{} allows users to take spreadsheets with minimal metadata, such as book title and author, and automatically 1) add authoritative, persistent identifiers like ISBNs 2) and cluster related \textit{Expressions} and \textit{Manifestations} of the same \textit{Work}, e.g., different translations or editions.
This enhancement makes it easier to combine related collections and analyze books at scale.
The tool is currently designed as an extension for OpenRefine—a popular software application—and connects to major bibliographic services including the Library of Congress, VIAF, OCLC, HathiTrust, Google Books, and Wikidata.
Our approach prioritizes human judgment.
Through an interactive interface, users can manually evaluate matches and define the contours of a \textit{Work}  (e.g., to include translations or not).
We evaluate reconciliation performance on datasets of U.S. prize-winning books and contemporary world fiction. 
\BR{} achieves near-perfect accuracy for U.S. works but lower performance for global texts, reflecting structural weaknesses in bibliographic infrastructures for non-English and global literature. 
Overall, \BR{} supports the reuse of bibliographic data across domains and applications, contributing to ongoing work in digital libraries and digital humanities.
\end{abstract}


\section{Introduction}

\begin{figure}
    \centering
    \includegraphics[width=1\linewidth]{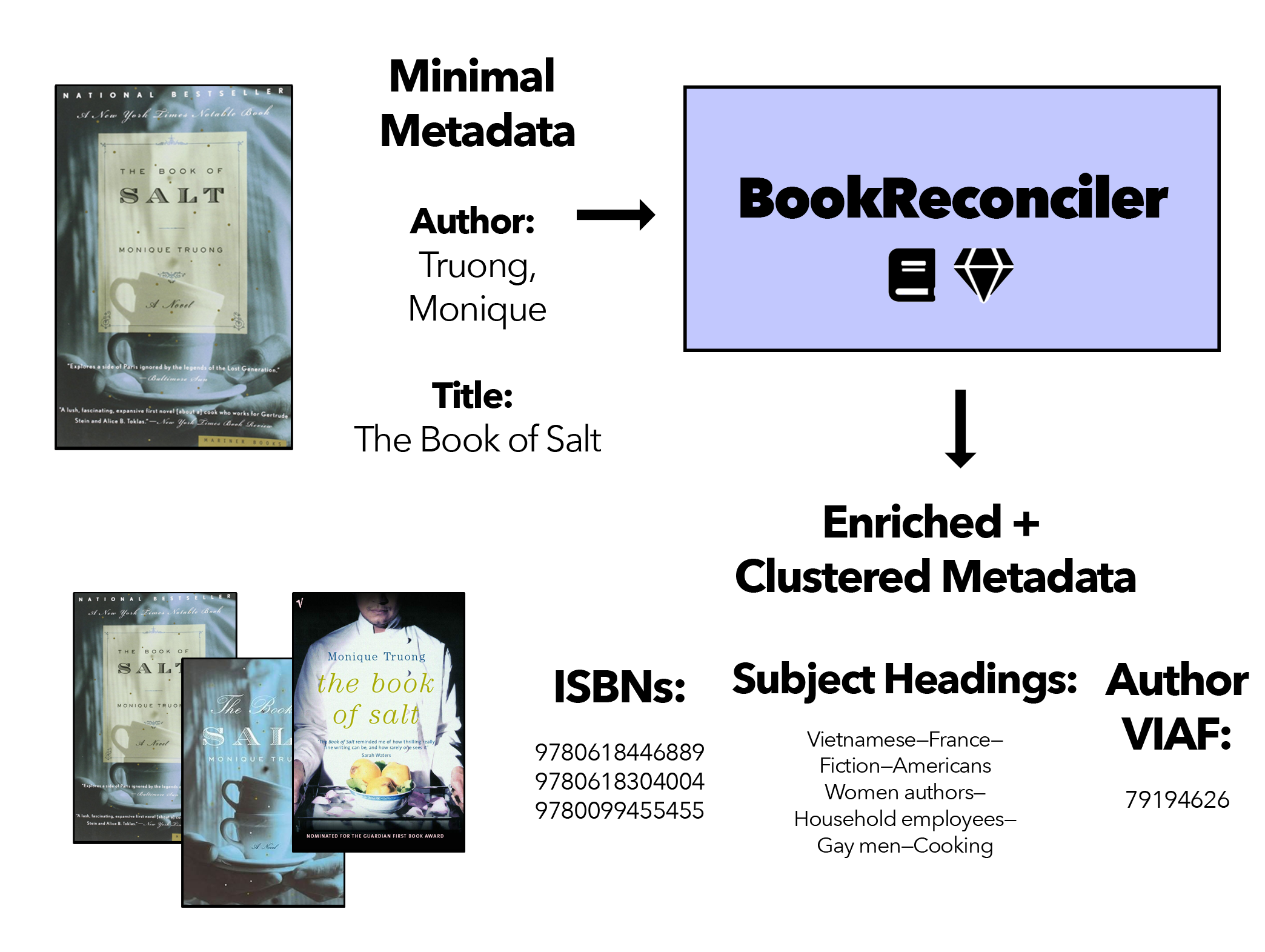}
    \caption{A conceptual demonstration of the \BR{} workflow. A user can submit a dataset with minimal bibliographic metadata, such as book title and author, and enrich the data with ISBNs, subject headings, VIAF identifiers, and more for related editions and formats—what we call a \textit{Work} cluster. The tool can be used to reconcile sources from the Library of Congress, Google Books, OCLC, HathiTrust, Wikidata, and VIAF.}
    \label{fig:logo}
\end{figure}

In many settings, people work with only minimal bibliographic metadata, often just a book’s title and author—for example, ``The Book of Salt'' by ``Monique Truong'' (Fig. \ref{fig:logo}).
Think of a humanities researcher compiling a list of prize-winning novels; an archivist stewarding an underdescribed collection; or a journalist assembling a dataset of banned books.

While basic information about these books may suffice for certain purposes, enriched metadata may be  necessary for others. For example, if users want to analyze genre or time period; connect to other data sources or library systems; or identify related editions, they will need to add subject headings, publications dates, persistent identifiers, and more. 
\textbf{What is the best way to enrich and cluster book metadata, especially at scale?}




This challenging question has come to the fore in the digital humanities, where researchers increasingly curate and publish bibliographic data, and where they often focus on books at the most abstract \textit{Work} level—in the sense of the Functional Requirements for Bibliographic Records (FRBR) model.
Examples include datasets of major U.S. literary prize winners \cite{grossman_index_2022}, bestselling novels \cite{pruett_new_2022, dileonardi_international_2025}, anthologies of African American literature \cite{earhartDALADatabaseAfrican2025}, and works of futuristic fiction \cite{wythoff_time_2025}.

Despite their great scholarly value, these sorts of datasets remain difficult to build upon because they often include minimal and inconsistent metadata. 
While on an individual basis we can see that ``The Book of Salt'' by ``Monique Truong'' refers to the same entity as ``The Book of Salt: A Novel'' by ``Truong, Monique,'' such discrepancies quickly become unwieldy at scale and cannot be resolved even by using computational text similarity approaches.

To address these challenges, we introduce \BR.\footnote{
\url{https://github.com/Post45-Data-Collective/openrefine-reconciliation-service}} Built as an extension for the widely used data-cleaning software application OpenRefine \cite{Delpeuch_Morris_Huynh_Weblate}, this tool supports bibliographic metadata enrichment and \textit{Work}-level clustering by drawing on six major bibliographic services: Library of Congress, VIAF, OCLC, HathiTrust, Google Books, and Wikidata. 
A user can take a spreadsheet with only book title and author information, and they can add persistent identifiers, as well as contextual information like genres, subject headings, publication dates, and descriptions.
They can also find and cluster different \textit{Manifestations} or \textit{Expressions} of the same \textit{Work} (e.g., translations, reprints, etc.).
The tool includes a human-in-the-loop interactive review interface that enables users to manually evaluate matches and customize a \textit{Work} according to their goals.

\BR{} has the potential to serve a wide community of users, including humanities researchers, librarians, archivists, journalists, and even the general public. 
By operating through OpenRefine's user-friendly, non-technical interface, we particularly seek to empower users with little to no computational expertise.

We evaluate BookReconciler on two datasets: U.S. prize-winning books and contemporary world fiction (representing 13 countries, nine languages, and five continents) \cite{grossman_index_2022, piper_mini_2025}. We find that the tool correctly matches 98\% of U.S. prize‐winners with Google Books, 99\% when using all services. For contemporary world fiction, the highest accuracy drops to 63\%. 
This suggests the tool is highly effective for well‐known English U.S. works, but performance is more variable for multilingual, geographically diverse literature.
Addressing these gaps in future work will require expanded coverage in major bibliographic services, as well as integration with additional international authority services.

\section{Background \& Related Work}

In the digital humanities, new data collectives and data-focused journals like the \href{https://data.post45.org/}{Post45 Data Collective} (with which we are affiliated), the \href{https://c19datacollective.com/}{Nineteenth Century Data Collective}, and the \href{https://openhumanitiesdata.metajnl.com/}{\textit{Journal of Open Humanities Data}} have begun to develop standards and guidelines for the creation and sharing of humanities data.
When working with bibliographic data, most recommend the inclusion of authoritative identifiers like ISBNs or HathiTrust Volume IDs.
However, this recommendation has been infrequently adopted by data creators, likely because Library and Information Science (LIS) practices and infrastructures have not been fully integrated into humanities research workflows.

Adding edition- or format-level identifiers like ISBNs is an important step that makes book data more interoperable.
But it still does not address larger issues related to representing and connecting books at the \textit{Work} level.
The concept of a \textit{Work} is a foundational principle in LIS. 
The most influential framework is the FRBR model, which introduced a four-tiered structure: \textit{Work}, \textit{Expression}, \textit{Manifestation}, and \textit{Item} \cite{tillett_what_2005, records_functional_1998}. 
For example, there are many editions, translations, and adaptations of a novel like Jane Austen’s \textit{Pride and Prejudice}, but these are considered \textit{Manifestations} or \textit{Expressions} of a single, overarching intellectual entity known as the \textit{Work}.

Scholars have pointed out that the FRBR model, while highly valuable and field-shaping, is not ``actionable''—that is, it does not provide specific guidelines about how this concept should be implemented technically \cite{coyleFRBRTwentyYears2015}. 
Major organizations like the Library of Congress and OCLC have struggled to implement the concept of a \textit{Work}, experimenting with various algorithmic clustering approaches and tools \cite{ bennett_concept_2003, vizine-goetz_classify_2010}.
This difficulty has led each system to develop their own definition, such as BIBFRAME Work entities \cite{library_of_congress_bibframe_nodate} or WorldCat Work IDs.
These inconsistencies complicate metadata reconciliation, particularly across heterogeneous sources.

\section{OpenRefine Background and Motivation}

The software application OpenRefine has been in active development since around 2008.
This web-browser based tool is popular with information professionals and researchers for data cleaning and manipulation.
The tool works on tabular data, allowing it to be inserted into a workflow as long as the dataset can be represented as a spreadsheet. 

OpenRefine also provides a mechanism to add reconciliation services that follow the W3C Reconciliation Service API standard. 
By designing \BR{} as an extension of OpenRefine, we can offer automated reconciliation with a non-technical, user-friendly interface that is familiar to many in the digital humanities, libraries, and elsewhere. 

\section{\BR{} Overview}

\begin{figure}
    \centering
    \includegraphics[width=1\linewidth]{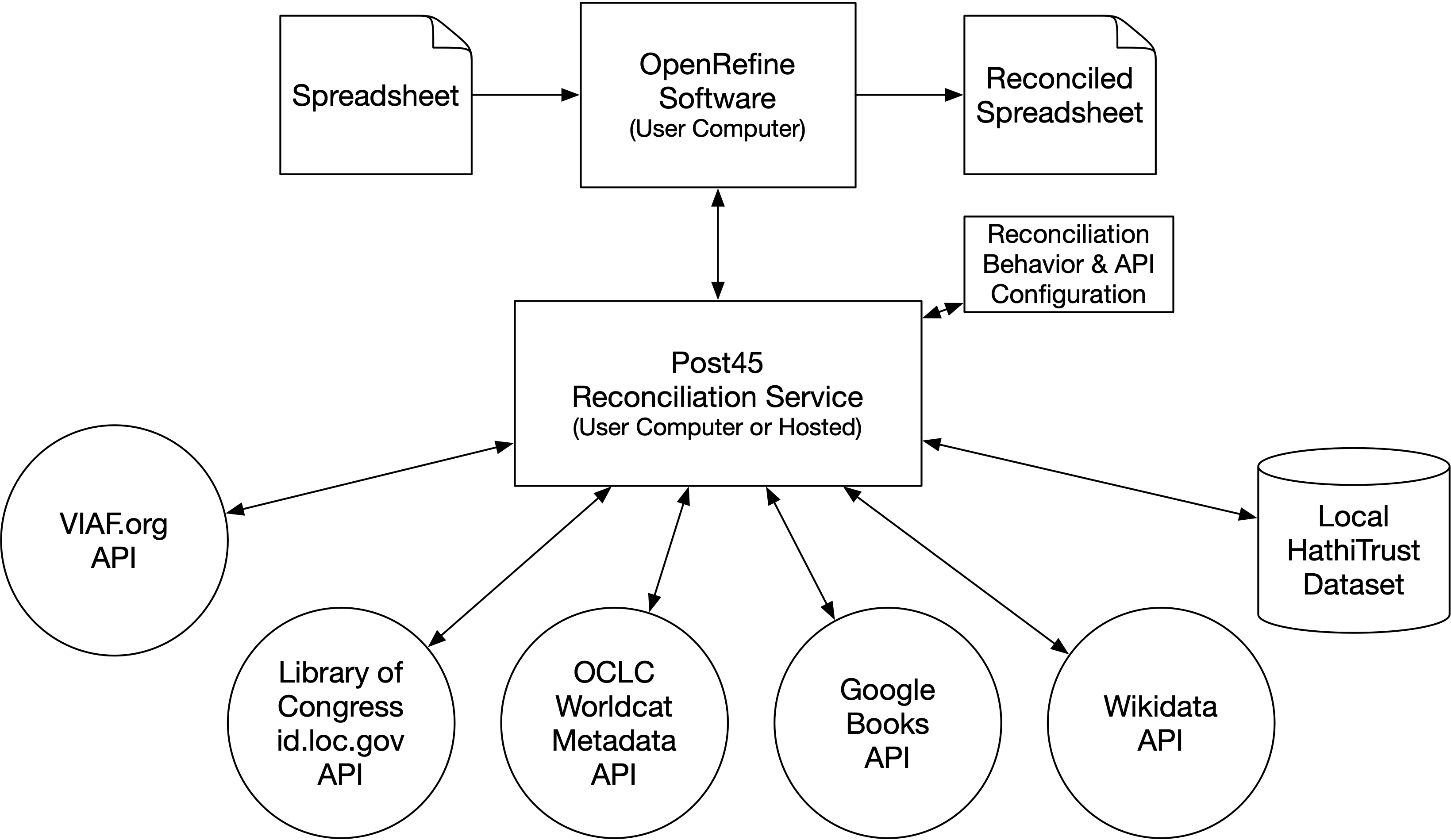}
    \caption{This diagram shows how the \BR{} tool, supported by the Post45 Data Collective, interacts with a user's spreadsheet, the OpenRefine software, and six distinct bibliographic services/sources: VIAF, Library of Congress, OCLC, Wikidata, and the HathiTrust Digital Library.}
    \label{fig:placeholder}
\end{figure}

When reconciling resources, it is best to cast a wide net. This improves the chances of correctly matching a resource and aggregating more identifiers.  
\BR{}\ supports data services including the Library of Congress, Google Books, VIAF, OCLC, Wikidata, and the HathiTrust Digital Library.
We select these services because they are among the most widely used, authoritative, and interoperable sources of book metadata available today.

These systems vary widely in the types of metadata they store and return, and in their access. 
We summarize key characteristics of the supported data services as follows:

\begin{itemize}
  \item \textbf{Library of Congress (id.loc.gov):} Public API access. Provides \textit{Work}-level search. \textit{Works} are narrowly scoped. Returns ISBN, LCCN, OCLC Numbers, LC \textit{Work} URI, and other metadata such as Subject Headings and Genres.

  \item \textbf{Google Books:} Public API access. Provides \textit{Manifestation}-level search. Returns ISBN and other metadata such as Description, Language, and Page Count.

  \item \textbf{VIAF (viaf.org):} Public API access. Provides cluster search for \textit{Works} (Name/Title) and Personal Names. \textit{Works} are broadly scoped. Returns VIAF \textit{Work} Identifiers.

  \item \textbf{OCLC WorldCat Metadata:} API key required. Provides \textit{Manifestation}-level search, but includes a \textit{Work} identifier to group related resources. Returns ISBN, OCLC numbers, LCCN, OCLC \textit{Work} IDs, Dewey (DDC), and other metadata such as Subject Headings, Genres, and Language.

  \item \textbf{Wikidata:} Public API and SPARQL endpoints provide \textit{Work}-level search. \textit{Works} are broadly scoped. Returns \textit{Work} IDs and links to external identifiers for enrichment.

  \item \textbf{HathiTrust:} No public API, but regular database dumps are available for local querying. Provides \textit{Manifestation}-level search. Returns ISBN, OCLC, LCCN, HathiTrust Volume IDs, and other metadata such as Earliest Publication Date, Latest Publication Date, and Thumbnail Image.
\end{itemize}

For all data services, reconciliation begins with a query—book title or author information—to return a matching result set. 
The tool attempts to cluster together resources, from the result set, that belong to the same \textit{Work} or author. 
Clustering is enabled by default, but users can configure the tool to reconcile only a single best match. 
This is useful in cases where precise matching is required, such as reconciling an exact list of publications from a specific year.

A key limitation of this tool is its reliance on external APIs. 
Since we do not have access to the full underlying databases, reconciliation is limited to the records returned in each API response.

\section{Architecture \& Workflow}

To begin reconciliation, the user first selects the column they wish to reconcile in OpenRefine—for example, the ``title'' column for books. 
Next, they select a reconciliation service, such as OCLC, Google Books, or HathiTrust. 
Then, they choose any additional columns—such as author/contributor name or publication date—to add as additional ``Properties,'' which can improve match accuracy. 
Finally, the user launches the reconciliation process with a single click.

\BR{} normalizes the submitted metadata and queries the selected API or data source to retrieve candidate matches. 
It ranks these potential matches using Levenshtein distance (tokenizing and alpha sorted), selecting the most likely result for each row.
The Levenshtein measurement is a ratio from 0 to 100, which can be customized by the user but is set at 80 by default.
If the service provides native \textit{Work} identifiers (as in the case of the OCLC WorldCat Metadata API), the tool uses the top-ranked result’s identifier to cluster together additional resources that share the same \textit{Work} value. 


\begin{figure}
    \centering
    \includegraphics[width=1\linewidth]{images/jcdl_paper_screen_shot_openrefine_recon.png}
    \caption{Additional metadata improves matches. This screenshot shows a user reconciling titles with HathiTrust and passing the additional column ``author'' (selected as ``Contributor Uncontrolled `First Last''') as an additional property. Users can pass additional metadata, such as author or book publication year, to increase accuracy of matches.}
    \label{fig:placeholder}
\end{figure}

The tool provides an interactive web interface that enables users to inspect and manually curate matches and \textit{Work} clusters. 
Users can hover over any reconciled cell to see a preview of matched resources, and can click to explore a more detailed view (in a separate Flask-based interface). 
For example, a user reconciling \textit{The Book of Salt} may choose to include or exclude translations from its \textit{Work} cluster, depending on their goals.
Additionally, users can navigate to the original source metadata, offering full transparency and provenance.


\begin{figure}
    \centering
    \includegraphics[width=1\linewidth]{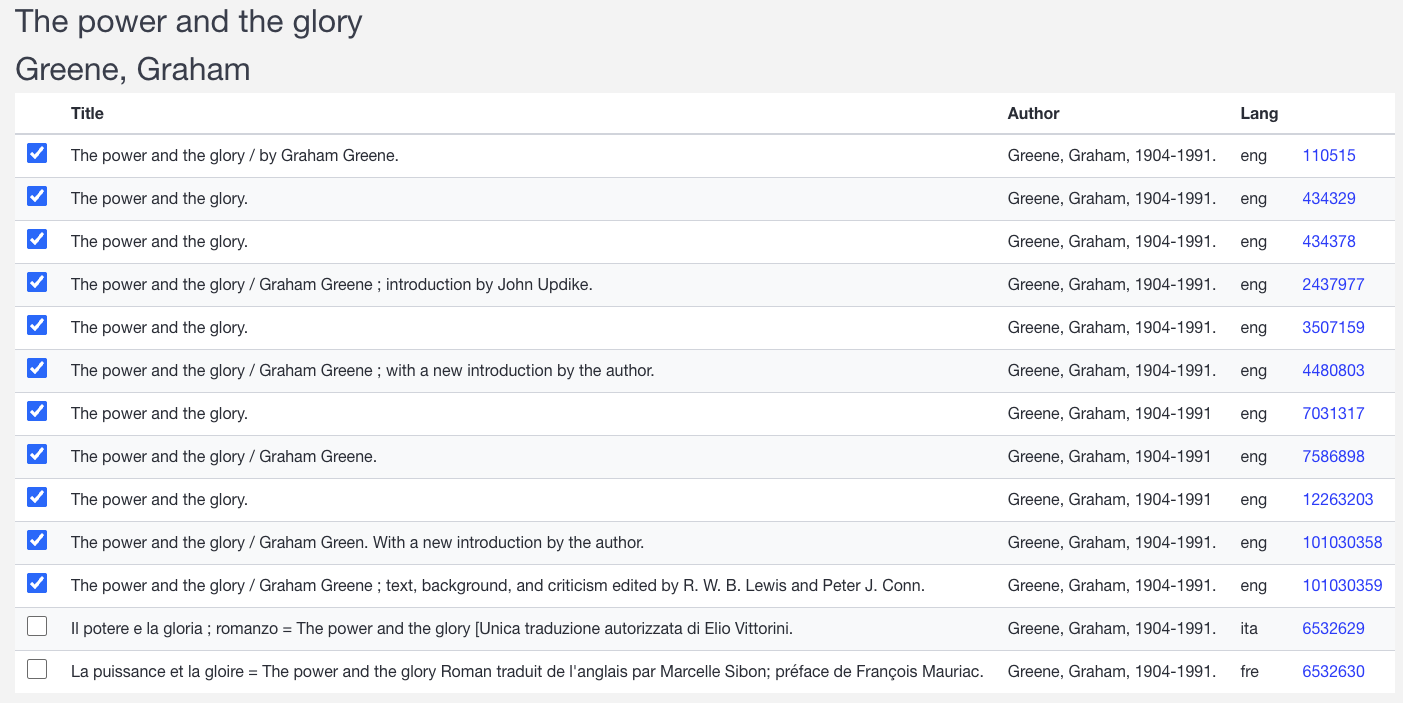}
    \caption{The interactive \textit{Work} cluster review interface. This screenshot shows the \textit{Work} cluster for Graham Greene's \textit{The Power and the Glory}. The user can select or de-select matched manifestations based on their goals. For example, they may want to include or exclude translations.}
            \label{fig:placeholder}

    \end{figure}

\begin{table*}[!t]
\centering
\caption{Reconciliation Accuracy with BookReconciler Across Bibliographic Services}
\label{tab:service-results}
\setlength{\tabcolsep}{4pt} 
\renewcommand{\arraystretch}{1.1}
\resizebox{\textwidth}{!}{%
\begin{tabular}{lcccccccccccccc}
\toprule
& Wikidata (Default)
& Library of Congress (BR)
& OCLC (BR)
& Google Books (BR)
& VIAF (BR)
& HathiTrust (BR)
& Wikidata (BR)
& All Services (BR) \\
\midrule
U.S Prize-Winning Books (1918-2020) & 36\% & 81\%  & 85\%  & \textbf{98\%} & 39\% & 57\% & 46\% & \textbf{99\%} \\
U.S Prize-Winning Books (Without Poetry)   & 51\% &  87\% &  88\%  & \textbf{99\%} &  51\%  & 60\% & 59\% & \textbf{99\%} \\
Contemporary World Fiction (2012-2023) & 4\% & 30\% & 36\%  & \textbf{63\%}  &  1\% & 0\% & 4\% & \textbf{63\%}  \\
\bottomrule
\end{tabular}%
}
\end{table*}

This workflow balances automation with human oversight. Users maintain control over how \textit{Works} are defined and clustered, which is particularly important given the diversity of bibliographic practices and the imperfections of even well-structured metadata systems.

Once reconciliation is complete, users can import additional fields such as ISBNs, genres, subject headings, or descriptions using OpenRefine’s ``Data Extension'' service. When fields contain multiple values (e.g., multiple ISBNs), the tool provides configuration options: values can be joined into a single cell with a delimiter, or exploded into multiple rows.

\section{Evaluation}

We evaluate BookReconciler on two datasets: books that won ``major'' (more than \$10,000) U.S. prizes between 1918-2020 (n = 691) \cite{grossman_index_2022} and contemporary world fiction published between 2012 and 2023 (n = 1,139) \cite{piper_mini_2025}.
The prize-winners include novels, poetry, as well as collections of essays and short stories.
Some of the books are now canonical, but others are much more obscure.
Poetry makes up 37\% of the total.
The world fiction draws from 13 countries and 9 languages.

We attempt to reconcile each book with each bibliographic service.
To provide a baseline comparison, we also test the general Wikidata reconciliation included in OpenRefine by default.
We pass in the title and full name of the author (not standardized) as an additional property.

For the U.S. dataset, BookReconciler correctly matches 98\% of titles with Google Books and 99\% when using all services. 
We find that performance degrades with poetry, and that variations in author name representation (e.g. “W.S. Merwin” vs “William Stanley Merwin”) can also degrade matching quality depending on the service.
On contemporary world literature, the highest accuracy (Google Books) drops to 63\%, with very low performance among other services. 
These results indicate that genre, metadata formatting, language, and region are significant contributing factors to reconciliation performance.

\section{Availability}

We release \BR{} under an open-source license (MIT), allowing researchers, librarians, and developers to freely use, adapt, and extend the tool.
The tool is currently available on GitHub \footnote{\url{https://github.com/Post45-Data-Collective/openrefine-reconciliation-service}} and can be installed as a one-click application or with Docker.
We also make available a \href{https://www.youtube.com/watch?v=V9ZJoFowRJM}{video tutorial and walk-through demonstration} \cite{mattmillerBookReconcilerDemoVideo2025}.
In the near term, maintenance will be supported by the Post45 Data Collective.
Long-term sustainability and new development will require broader community contributions and/or external funding.


\section{Conclusion \& Future Work}

We introduce BookReconciler, an open-source reconciliation tool that extends OpenRefine to support metadata enrichment and clustering. Our evaluation shows that the tool achieves near-perfect accuracy on U.S. prize literature (1918-2020) but performs less well on contemporary world literature (2012-2023).
Progress in this area will require major authority services to improve multilingual coverage.
The tool would also benefit from integrating additional international authority services, such as data.bnf.fr (France), Trove (Australia), NDL Linked Open Data (Japan), and others.  
Looking ahead, we see potential in using large language models as an additional layer to assess ambiguous matches, provided they are used thoughtfully and in combination with human judgment.


\bibliographystyle{IEEEtran}
\bibliography{ref}
\end{document}